\documentclass[twocolumn,english,prb,showpacs]{revtex4-1}
\usepackage{newcent}

\usepackage[T1]{fontenc}
\usepackage[latin9]{inputenc}
\usepackage{textcomp}
\usepackage{amsmath}
\usepackage{graphicx}

\makeatletter

\newcommand{\lyxmathsym}[1]{\ifmmode\begingroup\def\b@ld{bold}
  \text{\ifx\math@version\b@ld\bfseries\fi#1}\endgroup\else#1\fi}

\providecommand{\tabularnewline}{\\}

\@ifundefined{textcolor}{}
{%
 \definecolor{BLACK}{gray}{0}
 \definecolor{WHITE}{gray}{1}
 \definecolor{RED}{rgb}{1,0,0}
 \definecolor{GREEN}{rgb}{0,1,0}
 \definecolor{BLUE}{rgb}{0,0,1}
 \definecolor{CYAN}{cmyk}{1,0,0,0}
 \definecolor{MAGENTA}{cmyk}{0,1,0,0}
 \definecolor{YELLOW}{cmyk}{0,0,1,0}
 }

\@ifundefined{definecolor}
 {\@ifundefined{definecolor}
 {\usepackage{color}}{}
}{}

\usepackage{newcent}

\usepackage{textcomp}
\usepackage{relsize}

\DeclareRobustCommand{\lyxmathsym}[1]{\ifmmode\begingroup\def\b@ld{bold}
  \def\rmorbf##1{\ifx\math@version\b@ld\textbf{##1}\else\textrm{##1}\fi}
  \mathchoice{\hbox{\rmorbf{#1}}}{\hbox{\rmorbf{#1}}}
  {\hbox{\smaller[2]\rmorbf{#1}}}{\hbox{\smaller[3]\rmorbf{#1}}}
  \endgroup\else#1\fi}

\makeatother

\usepackage{babel}

\makeatother

\usepackage{babel}
\usepackage{pslatex}

\begin{document}

\title{Ultra-fast magnetisation rates within the Landau-Lifshitz-Bloch
model.\\
 }

\author{U. Atxitia}

\author{O. Chubykalo-Fesenko}

\affiliation{Instituto de Ciencia de Materiales de Madrid, CSIC, Cantoblanco,
28049 Madrid, Spain}

\begin{abstract}
 The ultra-fast magnetisation relaxation rates during the laser-induced magnetisation process are analyzed in terms of the Landau-Lifshitz-Bloch (LLB) equation for different values of spin $S$. The LLB equation is equivalent in the limit $S \rightarrow \infty$ to the atomistic Landau-Lifshitz-Gilbert (LLG) Langevin dynamics and for $S=1/2$ to the M3TM model [B. Koopmans, {\em et al.} Nature Mat. \textbf{9} (2010) 259]. Within the LLB model  the ultra-fast  demagnetisation time ($\tau_{M}$) and the transverse damping ($\alpha_{\perp}$) are parameterized by the intrinsic coupling-to-the-bath parameter $\lambda$, defined by  microscopic spin-flip rate. We show that for the phonon-mediated Elliott-Yafet mechanism,  $\lambda$ is proportional to the ratio between the non-equilibrium phonon and electron temperatures.
 We investigate the influence of the finite spin number and the scattering rate parameter $\lambda$ on the magnetisation relaxation rates. The relation between the fs demagnetisation rate  and the LLG damping, provided by the LLB theory, is checked basing on the available experimental data. A good agreement is obtained for Ni, Co and Gd favoring the idea that the same intrinsic scattering process is acting on the femtosecond and nanosecond timescale.

\end{abstract}


\maketitle

\section{Introduction}


Magnetisation precession and the spin-phonon relaxation rates at  picosecond
timescale  were considered to be the limiting factor for the speed of the
magnetisation switching \cite{NatureBack04,NatureTudosa04}, until using optical excitation with fs pulsed lasers
the possibility to influence the magnetisation on femtosecond timescale
was demonstrated \cite{BeaurepairePRL96,SchollPRL97,HohlfeldPRL97,KoopmansPRL00}. The
ultra-fast laser-induced demagnetisation immediately became a hot
topic of solid state physics due to an appealing possibility to push further
the limits of operation of magnetic devices. This ultra-fast process
has now been shown to proceed with several important characteristic
timescales \cite{KoopmansPRL00}: (i) the femtosecond demagnetisation with
 timescale $\tau_{M}$ (ii) the picosecond recovery
with timescale $\tau_{E}$ and (iii) the hundred picoseconds
-nanosecond magnetisation precession, traditionally characterized
by the ferromagnetic resonance frequency $\omega_{FMR}$ and the Landau-Lifshitz-Gilbert
damping parameter $\alpha_{LLG}$ (see Fig.1).

The physics of the magnetisation changes on femto-second timescales
is obviously not-trivial and will require novel theories within the
relativistic quantum electrodynamics of many electron systems.
 From theoretical point of view, the existing models try to answer an
open question of the role of different subsystems (photons, phonons,
electrons and spins) in the ultra-fast angular momentum  transfer \cite{StammNature07}.
This common goal is stimulated by experimental findings provided by the XMCD measurements
showing the important role  of the spin-orbit interactions \cite{BigotNaturePhys09, BoeglinNature10}.
For the present state of art quantum mechanical descriptions \cite{ZhangSpringer02,ZhangJAP08,ZhangPRL00,SteiaufPRB09,BigotNaturePhys09} of ultra-fast demagnetisation
processes involve unavoidable simplifications
and sometimes even some ad-hoc assumptions necessary to explain experimental findings, such as reduced exchange
interactions, enhanced spin-orbit coupling or a Gaussian
distribution of occupied states around the Fermi level.
While some degree of agreement has been achieved in modelling of the ultra-fast demagnetisation ($\tau_{M}$) scale \cite{KraussPRB09},
the modelling of all three ultra-fast
demagnetisation rates within the same approach is outside the
possibilities of the quantum mechanical approaches.

 The three-temperature (3T) phenomenological model involves the rate equations
 for the  electron, phonon and spin temperatures (energies) \cite{ZhangSpringer02,KaganovJETP57,AgranatZETF84,HohlfeldCP00}. Recently, it has been shown  that the introduction of the spin temperature is not adequate \cite{KazantsevaEPL08} since the spin system is not in the equilibrium on the femtosecond timescale. It has been suggested to couple  the spin dynamics to the two-temperature (2T) model for phonon and electron temperatures \cite{KazantsevaEPL08,Kazantseva07,KazantsevaPRB08,AtxitiaAPL07,AtxitiaPRB10}.  These models are based on the energy flow picture and leave unidentified the angular momentum transfer mechanism and the underlying quantum mechanism responsible for the spin flip \cite{AtxitiaPRB10}. They essentially
interpret the ultra-fast demagnetisation as "thermal" processes, understanding
the temperature as energy input from photon to electron and then to
the spin system.
By using these models the important role of the linear reversal path
in the femto-second demagnetisation has been identified \cite{KazantsevaEPL09,VahaplarPRL09}.
The comparison with experiment seems to indicate that in order to
have magnetisation switching in the ultra-fast timescale, a combined
action of "heat" and large field coming from the inverse Faraday effect
is necessary \cite{VahaplarPRL09}.

The most successful recent phenomenological models describing the ultra-fast
magnetisation dynamics are (i) the Langevin dynamics based on the
Landau-Lifshitz-Gilbert (LLG) equation and classical Heisenberg Hamiltonian
for localized atomic spin moments \cite{KazantsevaEPL08,Kazantseva07}, (ii) the Landau-Lifshitz-Bloch
(LLB) micromagnetics \cite{AtxitiaAPL07,AtxitiaPRB10} and (iii) the Koopmans's
magnetisation dynamics model (M3TM) \cite{KoopmansNatureMat09} .  The spin dynamics could be coupled to the
electron temperature from the 2T model, underlying the electronic
origin of the spin-flip process \cite{KazantsevaEPL08,Kazantseva07,AtxitiaAPL07,AtxitiaPRB10,VahaplarPRL09} or to both electron and phonon temperatures, underlying the Elliott-Yafet mechanism mediated by phonons \cite{KoopmansNatureMat09}. When the 2T model was carefully parameterised
from the measured reflectivity, it gave an excellent agreement with
the experiment in Ni \cite{AtxitiaPRB10} using the former approach or in Ni,  Co and Gd using the latter approach \cite{KoopmansNatureMat09}.

\begin{figure}
\includegraphics[angle=90,scale=0.35]{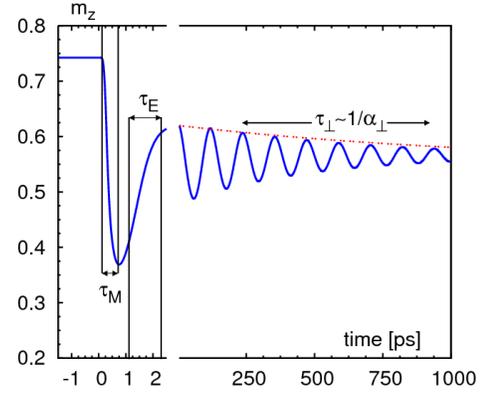}\caption{Characteristic time scales in  ultrafast laser-induced magnetisation
dynamics experiments. The curve is obtained by the integration of
the Landau-Lifshitz-Bloch equation coupled to the two-temperature model with the parameters from Ref.\onlinecite{AtxitiaAPL07}.
For the modelling of precession the applied field $H_{ap}=1$T at $30$ degrees was used.}
\label{LLB:graphs}

\end{figure}

In the classical derivation of the LLB equation the thermal averaging
has been performed analytically within the mean field (MFA) approximation
\cite{GaraninPRB97}. Thus, the LLB equation for classical spins ($S\rightarrow \infty$) is
equivalent to an ensemble of exchange-coupled atomistic spins  modelled by stochastic LLG equations \cite{KazantsevaPRB08,ChubykaloPRB06}.
At the same time, in some cases the LLB equation may be preferable
with respect to the atomistic Heisenberg model, since being micromagnetic
it can incorporate quantum nature of magnetism and the quantum derivation of LLB also exists \cite{GaraninQuantum}. In
particular the limits of validity for the statistical mechanics based on the classical Heisenberg model for
the description of materials with delocalized magnetism of \emph{d}-electrons in
transition metals or magnetism of \emph{f}-electrons in rare earths are not clear.
An alternative statistical simplified  description of \emph{d}-metals consists of a  two level system with spin-up and spin-down  bands (i.e. $S=\pm 1/2$), as has been done by B. Koopmans \emph{et al.} \cite{KoopmansNatureMat09}. Their model,
as we show in the present article, is also equivalent to the quantum LLB equation
with spin $S=1/2$.
An additional
advantage in the use of the LLB equation is the possibility to model larger spatial scales \cite{AtxitiaAPL07,KazantsevaPRB08}.
Therefore the LLB micromagnetics is an important paradigm
within the multiscale magnetisation dynamics description.
 The LLB equation has been shown to describe correctly the three stages
of the ultra-fast demagnetisation processes: the sub-picosecond demagnetisation,
the picosecond magnetisation recovery and the nanosecond magnetisation
precession \cite{AtxitiaAPL07,KazantsevaPRB08,AtxitiaPRB10}, see Fig.\ref{LLB:graphs}.

The intrinsic quantum mechanical mechanisms responsible for the ultra-fast demagnetisation
 in the LLB model are included in the intrinsic coupling-to-the-bath
parameter $\lambda$ \cite{GaraninQuantum,AtxitiaPRB10}. The coupling process
is defined by the rate of the spin flip.
 Several possible underlying quantum mechanisms are currently
under debate: the Elliott-Yafet (EY) electron scattering mediated by phonons or
impurities \cite{SteiaufPRB09,KoopmansNatureMat09}, or other electrons \cite{KraussPRB09} and electron-electron inelastic
exchange scattering \cite{HongPRB00,BalashovPRB08}.
 By combining the macroscopic
demagnetisation equation (M3TM model) with the rate of spin flip calculated on
the basis of full Hamiltonian, Koopmans \emph{et al.} \cite{KoopmansNatureMat09}
have been able to relate the ultra-fast demagnetisation time $\tau_M$ with the spin
flip rate of the phonon-mediated Elliott-Yafet scattering.  The authors fitted experimental demagnetisation rates in Ni, Co, Gd to the phenomenological M3TM model and  found them to be consistent with the values estimated on the basis of ab-initio theory \cite{SteiaufPRB09}.
The coupling-to-the-bath parameter $\lambda$ (microscopic damping parameter in atomistic LLG model) should be distinguished from that of the macroscopic damping $\alpha_{LLG}$ ($\alpha_{\perp}$ in the LLB model), a more complicated quantity which includes the magnon-magnon processes.

The first attempt to relate the sub-picosecond demagnetisation time with
the macroscopic damping processes was given by Koopmans \emph{et al.} \cite{KoopmansPRL00}
who suggested the relation $\tau_{M}\sim 1/\alpha_{LLG}$. Subsequently
and with the aim to check this relation several experiments in doped
permalloy were performed \cite{WalowskiPRL08,WoltersdorfPRL09,RaduPRL09}. The permalloy
thin films were doped with rare earth impurities, allowing to increase
in a controlled way the damping parameter $\alpha_{LLG}$. The effect
on the demagnetisation time $\tau_{M}$ was shown to be opposite \cite{RaduPRL09}
or null \cite{WalowskiPRL08}, in contrast to the above relation. However,
it should be noted that the analysis leading to this expression was
performed in terms of the Landau-Lifshitz-Gilbert equation,
relating the ultra-fast demagnetisation time $\tau_{M}$ to the transverse
damping without taking into account their temperature dependence.
Moreover, one should take into account that the rare-earth impurities may introduce
 a different scattering mechanism with a slower timescale \cite{WoltersdorfPRL09}.

Partially basing on the above mentioned  experimental results and from a general point of
view, the longitudinal relaxation (the ultra-fast demagnetisation rate
$\tau_{M}$) and the transverse relaxation (the LLG damping $\alpha_{LLG}$)
may be thought to be independent quantities. Indeed, different intrinsic and extrinsic mechanisms
can contribute to the demagnetisation rates at different timescales.
One can, for example, mention that during the femtosecond demagnetisation
the electron temperature is often raised up to the Curie temperature
\cite{AtxitiaPRB10,VahaplarPRL09}. At this moment, the high frequency THz spinwaves
\cite{BeaurepaireAPL04,DjordjevicPRB07} including the Stoner excitations \cite{BalashovPRB08}
contribute. At the same time, the transverse relaxation is related to
the homogeneous precessional mode. The LLB equation takes care of
 the different natures of longitudinal and
transverse relaxation, arising from the spin disordering. The LLB model calculates them independently but basing on
the same intrinsic scattering mechanism parameterized by the parameter $\lambda$.
The increment of the number of scattering events is mimicked by the increases of the electron temperature.
Consequently, the relation between the ultra-fast demagnetisation and precession remains valid but with a temperature-dependent correction.
 If this relation is confirmed
experimentally, a unique intrinsic coupling parameter means that the same main
microscopic mechanism is acting on both timescales. In the present
article we will show that the analysis of the available experimental data seems to indicate
towards this possibility, at least in pure transition metals such as Ni or Co and in rare earth metal Gd.
We did not find  validity of the corresponding relation in Fe.

 Up to now only
classical version ($S\rightarrow \infty$) of the LLB equation was used to model the ultra-fast
demagnetisation processes \cite{AtxitiaAPL07,KazantsevaPRB08,VahaplarPRL09}.
In the present article we show the important role of the choice of the quantum spin
value, resulting in the differences in the corresponding longitudinal
relaxation times. The article is organized as follows. In section II we present different formulations of the quantum LLB model and its main
features for different spin values $S$. In section III we present results on the modelling of the demagnetisation processes within LLB models with different choices of the quantum spins number $S$  and of the intrinsic scattering mechanisms. In section IV we present our attempts to link the ultra-fast demagnetisation rates in transition metals and Gd and comparison with available experimental data.
Section V concludes the article. In the Appendix to the article we demonstrate the equivalence of the LLB model with $S=1/2$ and the  M3TM model by B. Koopmans {\em et al.}\cite{KoopmansNatureMat09}.

\section{The Landau-Lifshitz-Bloch model with quantum spin number $S$.}

The LLB equation for a quantum spin was derived from the density matrix
approach \cite{GaraninQuantum}. Although the model Hamiltonian was rather
the simplest form of the spin-phonon interaction, the generalization
of the approach should be possible to more complex situations.  The macroscopic
equation for the magnetisation dynamics, valid at all temperatures,
is written in the following form:

\begin{equation}
\mathbf{\dot{n}}=\gamma[\mathbf{n}\times\mathbf{H}]+\frac{\gamma\alpha_{\parallel}}{n\lyxmathsym{\texttwosuperior}}\left[\mathbf{n}\cdot\mathbf{H}_{\textrm{eff}}\right]\mathbf{n}-\frac{\gamma\alpha_{\bot}}{n^{2}}\left[\mathbf{n}\times\left[\mathbf{n}\times\mathbf{H}_{\textrm{eff}}\right]\right]
\label{LLB}
\end{equation}
where $n=M/M_{e}(T)=m/m_{e}$ is the reduced magnetisation, normalized
to the equilibrium value $M_{e}$ at given temperature $T$ and $m=M/M_{e}(T=0K)$.
The effective field $\mathbf{H}_{\textrm{eff}}$, contains all usual micromagnetic contributions, denoted by $\mathbf{H}_{\text{int}}$
(Zeeman, anisotropy, exchange and magnetostatic) and is augmented
by the contribution coming from the temperature

\begin{equation}
\label{Field}
\mathbf{H}_{\textrm{eff}}=\mathbf{H}_{\textrm{int}}+\frac{m_{e}}{2\widetilde{\chi}_{\parallel}}\left(1-n^{2}\right)
\mathbf{n},
\end{equation}
where $\widetilde{\chi}_{\parallel}(T)=(\partial m /\partial H)_{H\rightarrow0}$ is the longitudinal susceptibility .
The LLB equation contains two relaxational parameters: transverse
$\alpha_{\bot}$ and longitudinal $\alpha_{\parallel}$, related to
the intrinsic coupling-to-the-bath parameter $\lambda$. In the quantum
description the coupling parameter $\lambda$ contains the matrix
elements representing the scattering events and, thus, is proportional
to the spin-flip rate due to the interaction with the environment.
This parameter, in turn, could be temperature dependent and, in our
opinion, it is this microscopic parameter which should be related to the Gilbert
parameter calculated through ab-initio calculations as in Refs.\cite{GilmorePRL99,KunesPRB02},
since the contribution coming from the spin disordering is not properly taken into account in these models.
In the quantum case the temperature dependence of the LLB damping parameters is given by the following expressions:

\begin{equation}
\alpha_{\parallel}=\frac{\lambda}{m_e}\frac{2T}{3T_{C}}\frac{2q_S}{\sinh\left(2q_S\right)}\underset{S\rightarrow\infty}{\Longrightarrow}  \frac{\lambda}{m_e}\frac{2T}{3T_{C}},
\label{longitud}
\end{equation}

\begin{equation}
\alpha_{\bot}=\frac{\lambda}{m_e}\left[\frac{\tanh\left(q_S\right)}{q_S}-\frac{T}{3T_{C}}\right]\underset{S\rightarrow\infty}{\Longrightarrow} \frac{\lambda}{m_e} \left[1-\frac{T}{3T_{C}}\right] ,
\label{transverse}
\end{equation}
with $q_S=3T_{C}m_{e}/[2(S+1)T]$, where $S$ is the quantum spin number
and $T_{C}$ is the Curie temperature. In the case $S\rightarrow\infty$ the damping coefficients have the
forms used in several previously published works \cite{GaraninChubykaloPRB04},
suitable for the comparison with the Langevin dynamics simulations
based on the classical Heisenberg Hamiltonian and in agreement with
them \cite{KazantsevaPRB08,ChubykaloPRB06}.

Eq.(\ref{LLB}) is singular for $T>T_{C}$, in this case it is more
convenient to use the LLB equation in terms of the variable $m=M/M_{e}(T=0K)$
\cite{ChubykaloPRB06}. The corresponding LLB equation is indistinguishable
from Eq.(\ref{LLB}) but with different relaxational parameters $\widetilde{\alpha}_{\parallel}=m_{e}\alpha_{\parallel}$,
$\widetilde{\alpha}_{\bot}=m_{e}\alpha_{\bot}$ and $\widetilde\alpha_{\bot}=\widetilde\alpha_{\parallel}$
for $T>T_{C}$, in this case the contribution of temperature to $\mathbf{H}_{\text{eff}}$ [the second term in Eq.(\ref{Field})] is $(-1 /\widetilde{\chi}_{\Vert })[1-3T_cm^{2}/5(T- T_c)m]
   \mathbf{m}$.  Although this formulation is more suitable
for the modelling of the laser-induced demagnetisation process, during
which the electronic temperature is usually raised higher than $T_{C}$,
it is the expression (\ref{transverse}) which should be compared with the
transverse relaxation parameter $\alpha_{LLG}$ due to the similarity of the formulation of the Eq.(\ref{LLB})
with the macromagnetic LLG equation. In the classical case and far from the Curie temperature $T \ll T_C$, $\lambda=\alpha_{\perp}=\widetilde{\alpha}_{\bot}$ ($\alpha_{LLG}$).

\begin{figure}[h!]
\includegraphics[scale=1]{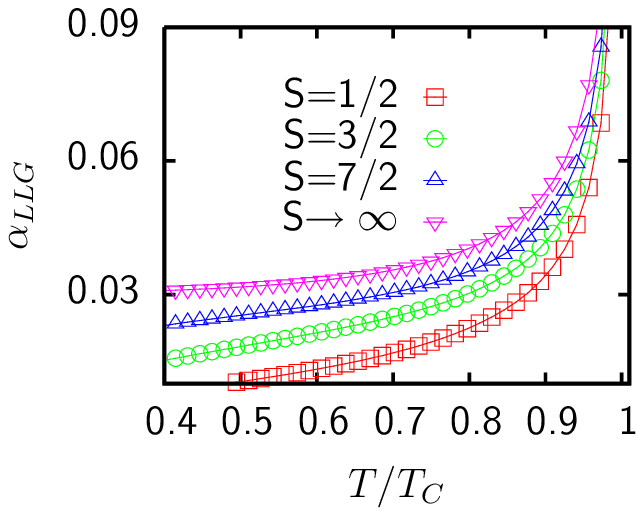}
\includegraphics[scale=1]{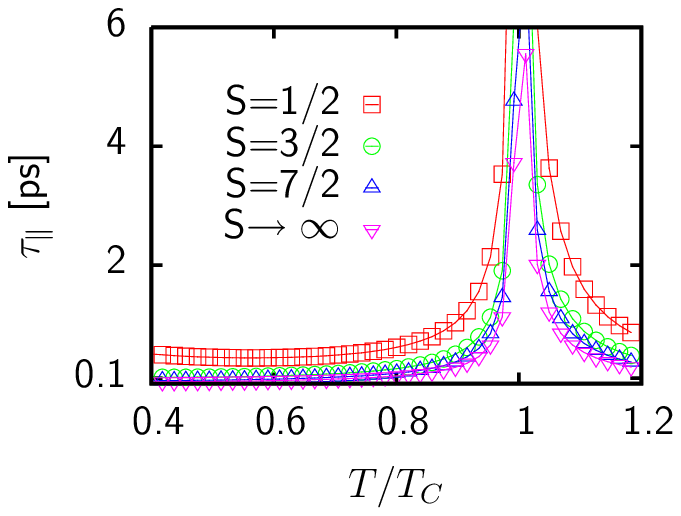}
\caption{ (Up) The transverse damping parameter $\alpha_{\perp}$ ($\alpha_{LLG}$)
as a function of temperature within the LLB model for different spin
values $S$. The intrinsic coupling parameter was set to $\lambda=0.03$. (Down) The longitudinal relaxation time $\tau_{\|}$ as a function of temperature
within the LLB model for different spin values $S$. The temperature-dependent magnetisation and the longitudinal susceptibility $\widetilde{\chi}_{\|}$  were evaluated in both cases in the MFA approach using the Brillouin function. }
\label{long}
\end{figure}

 In the "thermal" model  the nature of the longitudinal and the transverse relaxation differs from  the point of view of characteristic spinwave
frequencies. The transverse relaxation (known as the LLG damping) is basically the  relaxation
of the
 FMR mode. The contribution of other
spinwave modes is reduced to the thermal averaging of the micromagnetic
parameters and the main effect comes from the decrease of the magnetisation
at high temperature. Consequently, the transverse damping parameter
increases with temperature (see Fig.\ref{long}), consistent with atomistic
modelling results \cite{ChubykaloPRB06} and well-known FMR experiments
\cite{Bhagat,FarleJAP91}.

On the contrary, the main contribution to the longitudinal relaxation
comes from the  high-frequency spin waves. This process
occurs in a strong exchange field. As a result, the longitudinal relaxation
time (the inverse longitudinal relaxation) is much faster and increases with temperature,
known as critical slowing down, see Fig.\ref{long}. This slowing down has been
shown to be responsible for the slowing down of the femto-second demagnetisation
time $\tau_M$ as a function of laser pump fluency \cite{KazantsevaEPL08,AtxitiaPRB10}.
The characteristic longitudinal timescale is not only defined by the
longitudinal damping parameter (\ref{longitud}) but also by the temperature-dependent
longitudinal susceptibility $\widetilde{\chi}_{\|}(T)\cite{ChubykaloPRB06}$, according
to the following equation:
 \begin{equation}
\tau_{\|}(T)=\frac{\widetilde{\chi}_{\|}(T)}{\gamma\widetilde{\alpha}_{\|}(T)}.
\label{longtime}
\end{equation}

As it can be observed in Fig. \ref{long}  the transverse relaxation parameter  $\alpha_{\bot}$($\alpha_{LLG}$)
and the longitudinal relaxation time $\tau_{\|}$ have a strong dependence on the quantum spin number $S$ chosen to describe system's statistics. We conclude here about the occurrence of quite
different relaxation rates for the two extreme cases $S=1/2$
and $S=\infty$.

B. Koopmans {\em et al.} recently used a different equation to describe
the ultrafast demagnetisation dynamics \cite{KoopmansNatureMat09}, called M3TM model:
\begin{equation}
\frac{dm}{dt}=Rm\frac{T_{p}}{T_{C}}\left(1-m\coth\left(\frac{mT_{C}}{T_{e}}\right)\right).
\label{eq:KoopmansNature10}
\end{equation}
 Eq.(\ref{eq:KoopmansNature10}) has been obtained through the general Master equation approach for the dynamics of the populations of a two level system (spin $S=1/2$ was used) with
 the switching probability evaluated quantum-mechanically for the phonon-mediated EY spin-flips. Here
 $T_{p}$ and $T_e$ are phonon and electron temperatures, respectively, and $R$ is a material
specific parameter, related to the spin-flip probability in the phonon-mediated EY
scattering events $a_{\text{sf}}$, as
\begin{equation}
\label{R}
R=\frac{8 a_{\text{sf}} G_{ep}\mu_B k_B V_{a} T_C^2}{\mu_{\text{at}} E_D^2},
\end{equation}
where $V_a$ and $\mu_{\text{at}}$ are the atomic volume and magnetic moment, respectively,
$G_{ep}$ is the electron-phonon coupling constant, $k_B$ is the Boltzmann constant, $\mu_B$ is the Bohr magneton and $E_D$ is the Debye energy.
This equation has allowed to fit the ultra-fast demagnetisation time
($\tau_{M}$) obtaining the values of $R$ in Ni, Co and Gd \cite{KoopmansNatureMat09}
and relating them to the phonon-mediated EY scattering rates $a_{\text{sf}}$.

As we show in the Appendix, the M3TM equation (\ref{eq:KoopmansNature10}) corresponds to
the longitudinal part of the LLB equation with
thermal field only ($\mathbf{H}_{\textrm{int}}=0$) and with spin $S=1/2$, i.e. it is equivalent to

\begin{equation}
{dm \over dt} =\gamma \widetilde{\alpha}_{\|} H_{\text{eff}}.
\end{equation}
This gives a relation between
the intrinsic coupling parameter $\lambda$ and the  material
specific parameter $R$ and finally with the phonon-mediated EY spin-flip
probability $a_{\text{sf}}$ via the formula:

\begin{equation}
\lambda=\frac{3R}{2 \gamma}\frac{\mu_{\text{at}}}{k_{B}T_{C}}\frac{T_{p}}{T_{e}}=\lambda_0 \frac{T_{p}}{T_{e}}.
\label{lambda}
\end{equation}
Thus the two approaches are reconciled, provided that the temperature-dependent coupling
rate (\ref{lambda}) is used in the LLB equation, in contrast to other works \cite{KazantsevaEPL08,AtxitiaAPL07,AtxitiaPRB10} where the coupling
$\lambda$ is considered to be temperature-independent. Combining expressions (\ref{longtime}) (\ref{R}) and (\ref{lambda}),
one can immediately see that in the case of the phonon-mediated EY process, the longitudinal relaxation time is
  determined by
\begin{equation}
\tau_{\|}\propto \frac{\widetilde{\chi}_{\|}}{a_{\text{sf}}}\frac{E_D^2}{ G_{ep}V_a T_p}.
\label{longitudinaltime}
\end{equation}
In Ref.\onlinecite{KoopmansNatureMat09} and basing on the phonon-mediated EY picture, the classification of materials on the basis of the "magnetic interaction strength" parameter $\mu_{\text{at}}/ J$ was proposed, where $J$ is the material exchange parameter.
According to the expression above, the demagnetisation rate depends on more parameters, among which the important one is also the electron-phonon coupling $G_{ep}$ defining how fast the electron system can pass the energy to the phonon one. Another important parameter is the microscopic  spin-flip rate $a_{\text{sf}}$.
Comparing to the B. Koopmans \emph{et al.}\cite{KoopmansNatureMat09} materials classification, the longitudinal susceptibility in Eq.(\ref{longitudinaltime}) is indeed defined by the value of the atomic moment $\mu_{\text{at}}$ and by the fact that this function rapidly increases with temperature and
diverges close to  $T_C \propto J $.  At $T \approx T_C$ one  obtains a simple linear relation\cite{ChubykaloPRB06} $\widetilde{\chi}_{\|} \propto \mu_{\text{at}} /J$, thus showing the dependence of the demagnetisation rate on this parameter, as suggested in Ref.\onlinecite{KoopmansNatureMat09}.

 In the case of the phonon-mediated EY process the temperature dependence of the longitudinal relaxation is coming from the longitudinal susceptibiliy only (cf. Eq. (\ref{longitudinaltime})), as opposed to the case $\lambda=\text{const}$ (cf. Eq.(\ref{longtime})). (We do not discuss here the possibility that the phonon-mediated EY spin-flip rate $a_{\text{sf}}$ may be also temperature dependent.)  However, the temperature dependence of the susceptibility is characterized by its exponential divergence close to $T_C$. In these circumstances an additional linear temperature dependence  provided by the  longitudinal damping is  difficult to distinguish in the fitting procedure of experimental data.

\section{Modelling of the laser-induced ultra-fast demagnetisation within
the LLB models.}

In the spirit of Refs.\cite{KazantsevaEPL08,KazantsevaPRB08,AtxitiaPRB10,AtxitiaAPL07,KoopmansNatureMat09}
for the modelling of ultra-fast demagnetisation dynamics, the LLB
equation may be coupled to the electron temperature  $T_{e}$ only, understanding
the electrons as the main source for the spin-flip mechanism \cite{KazantsevaEPL08,KazantsevaPRB08,AtxitiaPRB10,AtxitiaAPL07} or to both phonon and electron temperatures in the spirit of the phonon-mediated Elliott-Yafet process  \cite{KoopmansNatureMat09}.
In both cases it is the electron temperature $T=T_{e}$ which couples to the magnetisation in the LLB formalism, since the phonon temperature  could only enter into the temperature dependence of the coupling-to-the bath parameter $\lambda$ via Eq.(\ref{lambda}) . Note
that the temperature $T$ is not the  spin temperature,
since the resulting dynamics is taking place out-of-equilibrium.

The electron $T_{e}$ and phonon $T_p$ temperatures are taken from the  two-temperature (2T)
model \cite{KaganovJETP57,Schoenlein87, AllenPRL87}. Within this model their dynamics is described by
two differential equations:
\begin{eqnarray}
C_{e}\frac{dT_{e}}{dt} & = & -G_{ep}(T_{e}-T_{p})+P(t),\nonumber \\
C_{p}\frac{dT_{p}}{dt} & = & G_{ep}(T_{e}-T_{p}).
\label{temp}
\end{eqnarray}
Here $C_{e}=\gamma_e T_e$ ($\gamma_e=\text{const}$)  and $C_{p}$ are the specific heats of the electrons
and the lattice. The Gaussian source term $P(t)$ is a function which
describes the laser power density absorbed in the material. The function
$P(t)$ is assumed to be proportional to the laser fluence $F$ with
the proportionality coefficient which could be obtained from the long time scale
demagnetization data (for which $T_{e}=T_{p}$) \cite{AtxitiaPRB10}. The dynamics of the electron temperature can be also
measured directly in the time-resolved photoemission experiment \cite{BovensiepenJPCM07}.

 The first of Eqs.(\ref{temp})
may also include a diffusion term $\nabla_z(\kappa \nabla_z T_e)$ taking into account a final penetration
depth of the deposited energy into the film thickness \cite{KoopmansNatureMat09} and a term, $C_e (T_e-300 K)/\tau_{th}$ describing the heat
diffusion to the external space \cite{AtxitiaPRB10}.  In the present article, the parameters
for the 2T-model were taken either  from Koopmans {\em et al.}
\cite{KoopmansNatureMat09} or from U. Atxitia {\em et al.} \cite{AtxitiaPRB10} (for Ni only),  where they were carefully parameterized through the reflectivity measurements. The
Ni (Co, Gd etc) parameters, such as magnetisation as a function of temperature
were taken assuming the Brilloiun (Langevin for $S \rightarrow \infty$) function.

The coupling of the 2T model to the LLB
equation adequately describes all three stages of the ultra-fast demagnetisation
rates: sub-ps demagnetisation, ps recovery and sub-ns precession \cite{AtxitiaAPL07, AtxitiaPRB10},
see Fig.1. As a consequence of the temperature dependence of both longitudinal
damping and susceptibility, and since the temperature is dynamically
changed according to Eqs.(\ref{temp}), the longitudinal relaxation
time is time-dependent via Eq.(\ref{longtime}). It is also strongly dependent
on the parameters of the 2T model and its dynamics is not simple. Consequently, the sub-ps ultra-fast demagnetisation
generally speaking is not exponential and cannot be described in terms
of one relaxation time $\tau_{M}$. Simple analytical expression
is possible to obtain with the supposition of a square-shaped temperature
pulse \cite{KazantsevaEPL09}. The two-exponential fitting is also often used \cite{DjordjevicPRB07,AtxitiaPRB10}. In our approach the fs demagnetisation  is fitted directly to the solution of the LLB equation without assumption of the one- or two-exponential decay. However, to comply with the existing approaches, we still discuss the demagnetisation rate in terms of a unique parameter $\tau_M$.

\begin{figure}[h!]
\includegraphics[scale=1.0]{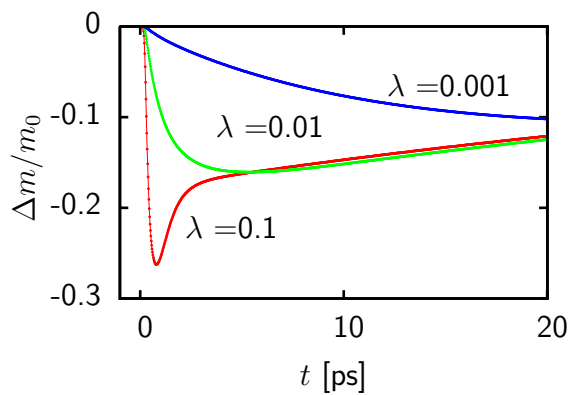}
\caption{The result of integration of the LLB model ($S\rightarrow \infty$) with different parameters $\lambda$ (increasing from top to the bottom). In this case the the 2T model parameters were taken from Ref.\onlinecite{AtxitiaPRB10} with laser fluence $F=30$ mJ/cm$^2$  }
\label{diversity}
\end{figure}

In the experiment performed in the same material the only remaining fitting parameter for the LLB model is the coupling parameter $\lambda$.
The choice of $\lambda$ together with the parameters of the 2T model  defines  all magnetisation  rates. In Fig.\ref{diversity} we present modelling of the ultra-fast demagnetisation and remagnetisation for various values of the coupling parameter $\lambda$, chosen to be independent on temperature, as in Ref. \onlinecite{AtxitiaPRB10}. If for some reason the scattering channel was suppressed, this would lead to a small scattering rate and consequently a small demagnetisation and a slow recovery. Indeed,  the value of $\lambda$ for Gd was found to be 60 times smaller than for Ni (see Table I). This small value of $\lambda$ assures a large delay in the magnetisarion relaxation towards the equilibrium electron temperature.  Thus this parameter defines the diversity of the demagnetisation rates in larger extend than the ratio $\mu_{\text{at}}/J$, suggested in  Ref.\onlinecite{KoopmansNatureMat09} and discussed in the previous subsection.

\begin{figure}[h!]
\includegraphics[scale=1.0]{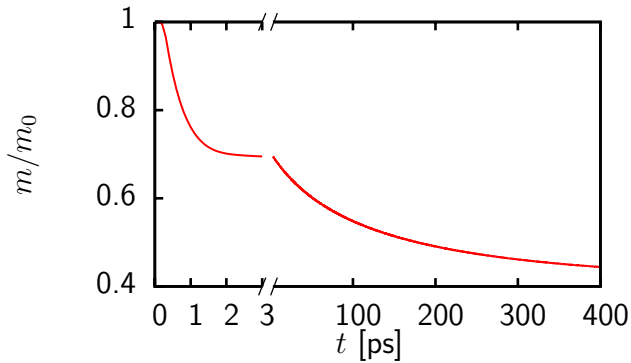}
\caption{The result of integration of the LLB model ($S \rightarrow \infty$) with constant $\lambda_0=0.0015$ (see Table I). In this case the  2T model parameters were taken from Ref.\onlinecite{KoopmansNatureMat09} corresponding to Gd. }
\label{small_l}
\end{figure}

Another parameter strongly influencing the demagnetisation rates is the phonon-electron coupling $G_{ep}$ defining the rate of the electron temperature equilibration time. This is the main parameter governing the magnetisation recovering time $\tau_E$. Indeed, in Ref.\cite{KoopmansNatureMat09} the phonon-electron coupling $G_{ep}$ was chosen to be 20 times smaller for Gd than for Ni.
By adjusting this parameter, the ultra-slow demagnetisation rates observed in TbFe alloy \cite{TbFe}, Gd\cite{Wietstruk} and in half-metals \cite{MullerNatureMat09} as well as the two time-scales demagnetisation \cite{Wietstruk,KoopmansNatureMat09} are also well-reproduced (see, as an example, Fig.\ref{small_l}). Within this model the two-time scale process consists of a relatively fast demagnetisation (however much slower than in Ni), defined by the electron temperature and small value of $\lambda$, followed by a much slower process due to a slow energy transfer from the electron to the lattice system.

\begin{figure}[h!]
\includegraphics[scale=1.0]{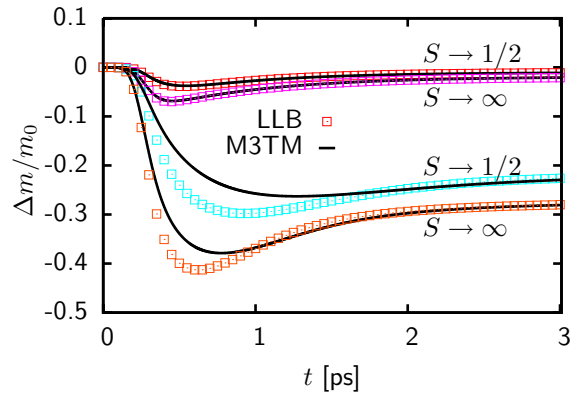}

\caption{Magnetisation dynamics during laser-induced demagnetisation process
calculated within the LLB model with different spin numbers and for two laser-fluencies
$F=10$ mJ/cm$^2$ (upper curves) and $F=40$ mJ/cm$^2$ (bottom curves). Ni parameters from Ref.\onlinecite{AtxitiaPRB10} were used. The symbols are calculated
with the LLB equation with the intrinsic damping parameter
using a constant $\lambda_0=0.003$ value, and the solid lines with the LLB equation
and the intrinsic coupling with the temperature dependent $\lambda=\lambda_0\left(T_{p}/T_{e}\right)$.}

\label{pumpprobedemag}
\end{figure}

As it was mentioned in the previous subsection, the phonon-mediated EY mechanism  predicts  the coupling to the bath parameter $\lambda$ to be
dependent on the ratio between the phonon and electron temperature through the relation (\ref{lambda}). A decrease of $\lambda$ up to two times
at high fluencies is observed for Ni and Co. The analysis of the data presented in Ref. \onlinecite{KoopmansNatureMat09} and \onlinecite{BovensiepenJPCM07} for Gd  has shown that during the demagnetisation process the ratio $T_e/T_p$ has increased almost 6 times. In Fig.\ref{pumpprobedemag} we present the magnetisation dynamics for Ni evaluated for two
laser pulse fluencies, assuming various values of the spin $S$  and temperature-dependent and independent $\lambda$ values. Note quite different demagnetisation
rates at high fluency for two limiting cases $S=1/2$, used in Ref.\onlinecite{KoopmansNatureMat09}
and $S=\infty$, used in Ref.\onlinecite{AtxitiaPRB10}.
 The differences in the choice of $\lambda$ are pronounced at high pump fluency but are not seen at low fluency.
 One can also hope that in the fitting procedure of experimental data it would be possible to distinguish the two situations.
Unfortunately, the fitting to experimental data procedure is complicated and the changes coming from the two cases described above are competing with several different possibilities such as an additional temperature dependency in electron or phonon specific heats \cite{WangPRB10}. Additionally, we would like to mention different electron-phonon coupling constants $G_{ep}$ used in Refs. \onlinecite{AtxitiaPRB10} and \onlinecite{ KoopmansNatureMat09}.  Fitting to experimental data from Ref.\onlinecite{KoopmansNatureMat09} for Ni for high fluence, we have found that the case of the temperature-dependent $\lambda=\lambda_0 (T_p /T_e)$ can be equally fitted with the temperature-independent $\lambda \approx  \lambda_0/2$. To answer definitely which fitting is better, more experimental data promoting one or another intrinsic mechanism and varying laser fluency is necessary.

\section{Linking different timescales}

Since the longitudinal relaxation occurs under strong exchange field and the transverse relaxation - under external applied field, their characteristic timescales are quite different. However,
the LLB equation  provides a  relation between the ultra-fast demagnetisation
(longitudinal relaxation) and the transverse relaxation (ordinary
LLG damping parameter) via the parameter $\lambda_0$ ($\lambda=\lambda_0$ or $\lambda=\lambda_0 (T_ p/T_e)$ for $T_p=T_e$).
The two demagnetisation
rates could be measured independently by means of the ultra-fast laser
pump-probe technique \cite{WalowskiJPhysD}. It has been recently demonstrated
\cite{Mizukami} that the damping of the laser-induced precession coincides
with the measured by FMR in transition metals. By separate measurements
of the two magnetisation rates, the relations (\ref{transverse}) and (\ref{longtime}) given
by the LLB theory could be checked. This can provide the validation of the LLB model, as well as the answer to the question if the same
microscopic mechanism is acting on femtosecond and picosecond timescales. Unfortunately, the damping problem in ferromagnetic materials
is very complicated and the literature reveals the diversity of measured values in the same material, depending on the preparation conditions.

Thus, to have a definite answer the measurement on the same sample is highly desired. The measurements of both $\alpha_{\bot}$ and $\tau_M$ are available for Ni \cite{AtxitiaPRB10}
where an excellent agreement between ultra-fast magnetisation rates  via a unique temperature-independent parameter $\lambda=0.04$ has been reported \cite{AtxitiaPRB10}. The results of the systematic measurements of $\tau_M$ are also available for Ni, Co, Gd in Ref. \onlinecite{KoopmansNatureMat09}, as well as for Fe \cite{CarpenePRB08}. The next problem which we encounter here is that the demagnetisation rates strongly depend on the spin value $S$, as is indicated in Figs. \ref{long} and \ref{pumpprobedemag}. The fitting of experimental data using LLB model with different $S$ values results in different values of the coupling parameter $\lambda_0$. The  use of $S=1/2$ value \cite{KoopmansNatureMat09} or $S=\infty$ value \cite{AtxitiaPRB10} is quite arbitrary and these values do not coincide with the atomic spin numbers of Ni,Co, Gd.
 Generally speaking, for  metals the spin value is not a good quantum number. The measured temperature dependence of magnetisation, however, is well fitted by the Brillouin function with $S=1/2$ for Ni and Co and $S=7/2$ for Gd \cite{Cullity}. These are the values of $S$ which we use in Table I.

 Consequently in Table I we present data for the coupling parameter $\lambda_0$ extracted from Ref.\onlinecite{ KoopmansNatureMat09}. Differently to this article, for Gd we corrected the value of the parameter $R$ to account for a different spin value by the ratio of the factors, i.e.  $R^{S_1}=\left(f_{S_2}/f_{S_1}\right)R^{S_2}$ with

\begin{equation}
f_S=\frac{2q_{S}}{\sinh\left(2q_{S}\right)}\frac{1}{m_{e,S}^{2}\chi_{\parallel}^{S}},
\end{equation}
where the parameters are evaluated at $120 K$ using the MFA expressions for each spin value $S$.
The data are evaluated for the phonon-mediated EY process  with the temperature-dependent parameter
$\lambda$ via the expression (\ref{lambda}).
The value of the Gilbert damping parameter $\alpha_{\perp}$ was then estimated through formula (\ref{transverse}) at $300 K$ (for Ni and Co) and at $120 K$ for Gd. Note that for temperature-independent $\lambda=\lambda_0$ the resulting $\lambda_0$ and $\alpha_{\perp}$ values are approximately two times smaller for Ni and Co. The last column presents experimental values for the same parameter found in literature for comparison with the ones in the fifth column, estimated through measurements of the ultra-fast demagnetisation times $\tau_M$ and the relation provided by the LLB equation.

\begin{table}
\begin{tabular}{ccccccc}
\hline
Material &$S$ & $R\cite{KoopmansNatureMat09}$ & $\lambda_0$&   & $\alpha_{\perp}$  & $\alpha_{\rm LLG}$ \tabularnewline
\hline
\hline
Ni  & $1/2$     & $17.2$ & $0.0974$      &  & $0.032$ & $0.019$\cite{Dewar}-$0.028$ \cite{Bhagat}\tabularnewline
Co  &$1/2$     & $25.3$  & $0.179$       &  & $0.025$ & $0.0036$\cite{Bhagat}-$0.006$\cite{Lindner}-$0.011$ \cite{Heinrich}\tabularnewline
Gd   &$7/2$     & $ 0.009$ & $0.0015$       &   & $0.00036$ & $0.0005$ \cite{WoltersdorfPRL09}\tabularnewline

\end{tabular}

\caption{The data for ultra-fast demagnetisation rate parameters for three different metals
from ultrafast demagnetization rates and from FMR mesurements. The third column presents the demagnetisation
parameter $R$ from Ref. \onlinecite{KoopmansNatureMat09}, corrected in the case of Gd for spin $S=7/2$. The fourth column presents the value of the $\lambda_0$ parameter, as estimated from the M3TM model \cite{KoopmansNatureMat09} and the formula Eq.(\ref{lambda}). The fifth column presents the  data for $\alpha_{\perp}$ estimated via the LLB model Eq.(\ref{transverse}) and the $\lambda_0$ value from the third column, at room temperature $T=300 K$ for Co and Ni and at $T=120K$ for  Gd .
The last column presents the experimentally measured Gilbert damping collected from different references.}
\label{table}
\end{table}

Given the complexity of the problem, the results presented in Table I demonstrate quite a satisfactory agreement between
the values, extracted from the ultra-fast demagnetisation time $\tau_{M}$ and the Gilbert damping
parameter $\alpha_{\perp}$ via one unique coupling-to-the-bath parameter
$\lambda$. The agreement is particularly good for Ni, indicating that the same spin flip mechanism is acting
on both timescales. This is true for both experiments in Refs.\onlinecite{AtxitiaPRB10} and \onlinecite{KoopmansNatureMat09}.
For Co the value is some larger. For the temperature-independent $\lambda$, the resulting value is two times smaller and the agreement is again satisfactory. We would like to note that no good agreement was obtained for Fe. The reported damping values \cite{Bhagat} are 5-10 times  smaller as estimated from the demagnetisation rates measured in Ref. \onlinecite{CarpenePRB08}.

\section{Conclusions}

The Landau-Lifshitz-Bloch (LLB) equation provides a micromagnetic tool for the phenomenological modelling of the ultra-fast demagnetisation processes. Within this model one can describe the temperature-dependent magnetisation dynamics at arbitrary temperature, including close and above the Curie temperature. The micromagnetic formulation can take into account the quantum spin number. The LLB model includes the dynamics governed  by both the atomistic LLG model and the M3TM model by Koopmans \emph{et al.}\cite{KoopmansNatureMat09}. In the future it represents a real possibility for the multiscale modelling \cite{KazantsevaPRB08}.

We have shown that within this model the ultra-fast demagnetisation rates could be parameterized through a unique temperature dependent or independent parameter $\lambda$, defined by the intrinsic spin-flip rate.  The magnetisation dynamics is coupled to the electron temperature through this parameter and is always delayed in time. The observed delay is higher for higher electron temperature. This is in agreement with the experimental observation that different materials demagnetize at different rates \cite{KoopmansNatureMat09, MullerNatureMat09} and that the process is slowed down with the increase of laser fluency. We have shown that for the phonon-mediated EY mechanism the intrinsic parameter $\lambda$ is dependent on the ratio between phonon and electron temperatures and therefore is temperature dependent on the femto second - several picosecond timescale. The LLB equation can reproduce slow demagnetizing rates observed in several materials such as  Gd, TbFe and half metals. This is in agreement with both phonon-mediated EY picture since in Gd a lower spin-flip probability was predicted  and also with the inelastic electron scattering picture, since the electron diffusive processes are suppressed in insulators and half-metals \cite{BattiatoPRL10, MullerNatureMat09}. However, we also stress the importance of other parameters determining the ultra-fast demagnetisation rates, such as the electron-lattice coupling.

The macroscopic damping parameters (longitudinal and transverse) have different natures in terms of the involved spinwaves and in terms of the timescales. Their temperature dependence is different, however, they are related by the spin-flip rate.
We have tried to check this relation in several transition metals such as Ni, Co, Fe and the rare-earth metal Gd.
A good agreement is obtained in  Co and Gd and an excellent agreement in Ni. This indicates that on both timescales the same main microscopic mechanism is acting. In Ni the agreement is good  both within  the assumptions $\lambda=\lambda_0$ and $\lambda=\lambda_0 T_ p/T_e$. In Co  the agreement seems to be better with the temperature-independent parameter $\lambda=\lambda_0$ which does not indicate towards the phonon-mediated EY mechanism. However, given a small discrepancy and the complexity of the damping problem, this conclusion cannot be considered definite.  We can neither exclude an additional temperature dependence of the intrinsic scattering probability (i.e. the parameter $\lambda_0$) for both phonon-mediated EY and exchange scattering mechanisms which was not taken into account.

 An open question is the problem of doped permalloy where an attempt to systematically change the damping parameter by doping with rare-earth impurities was undertaken \cite{WoltersdorfPRL09} in order to clarify the relation between the LLG damping and the ultra-fast demagnetisation rate \cite{WalowskiPRL08,RaduPRL09}.
The results are not in agreement with the LLB model. However in this case we think that the hypothesis of the slow relaxing impurities presented in Ref.\onlinecite{RaduPRL09} might be a plausible explanation. Indeed, if the relaxation time of the rare earth impurities is high, the standard LLB model is not valid since it assumes an uncorrelated thermal bath.
The correlation time could be  introduced in the classical spin dynamics via the Landau-Lifshitz-Miyasaki-Seki approach \cite{AtxitiaPRL09}. It has been shown that the correlation time of the order of 10 fs slows down the longitudinal relaxation independently on the transverse relaxation. Thus in this case, the modification of the original LLB model to account for the colored noise is necessary.

\section{Acknowledgement}

This work was supported by the Spanish projects MAT2007-66719-C03-01,
CS2008-023.

\appendix
\section{}
To show the equivalence between the LLB model with $S=1/2$ and the M3TM model \cite{KoopmansNatureMat09},
we compare the relaxation rates resulting from both equations.
We start with the M3TM equation

\begin{equation}
\frac{dm}{dt}=-R\frac{T_{p}}{T_{C}}\left(1-m\coth\left[\left(\frac{T_{C}}{T_e}\right)m\right]\right)m
\label{M3TM_A}
\end{equation}
where we identify the Brillouin function for the case
$S=1/2$, $B_{1/2}=\tanh\left(q\right)$ with $q=q_{1/2}=\left(T_{C}/T_e\right)m$. Now, we use the identity $B_{1/2}=2/B'_{1/2}\sinh\left(2q\right)$ to write
\begin{equation}
\frac{dm}{dt} =  -R\frac{T_p}{T_{C}}\left[\frac{2}{\sinh\left(2q\right)}\right]\left(\frac{1-\frac{B_{1/2}}{m}}
 {B'_{1/2}}\right)m^2
 \end{equation}
 we multiply and divide by $q \mu_{\text{at}}\beta$ to obtain
\begin{equation}
\frac{dm}{dt} =  -R\frac{T_p}{T_{C}} \frac{\mu_{at}}{k_B T_C} \left[\frac{2q}{\sinh\left(2q\right)}\right]\left(\frac{1-\frac{B_{1/2}}{m}}
 {\mu_{\text{at}}\beta B'_{1/2}}\right)m
 \label{koopmansmanipulate}
 \end{equation}
\begin{figure}[h!]
\includegraphics[scale=1.0]{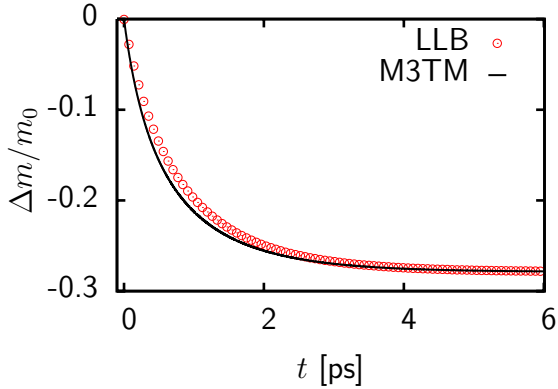}
\caption{Longitudinal relaxation calculated with M3TM and LLB (S=1/2) models for Nickel parameters \cite{AtxitiaPRB10} and $T/Tc=0.8$.}
\label{compare}
\end{figure} 
 
We expand around equilibrium $m_{e}=B_{1/2}(q_{e})$ the  small quantity $1-B_{1/2}/m$
\begin{equation}
1-\frac{B_{1/2}(q)}{m}  \cong  \frac{\delta m}{m_e}\left(1-\left(\frac{T_C}{T_e}\right)B_{1/2}'(q_{e})\right)
\end{equation}
where $\delta m=m-m_{e}$ . Next, we expand $m$ around $m^2_{e}$
\begin{equation}
m=m_e+\frac{1}{2}\frac{(m^2-m_e^2)}{m_e} \Longrightarrow \frac{\delta m}{m_e} =\frac{(m^2-m_e^2)}{2m_e^2}
\end{equation}
and,
\begin{equation}
\frac{1-B_{1/2}/m}{\beta\mu_{\text{at}} B_{1/2}'}\approx\frac{1}{2\widetilde{\chi}_{\|}}\frac{(m^{2}-m_{e}^{2})}{m_{e}^{2}}
\label{susapprox}
\end{equation}
Finally, collecting the equations (\ref{koopmansmanipulate}) and (\ref{susapprox}) altogether:
\begin{equation}
\frac{dm}{dt}=\left(\frac{3R}{2}\frac{\mu_{\text{at}}}{k_{B}T_{C}}\right)\frac{2T_p}{3T_{C}}\frac{2q}{\sinh\left(2q\right)}
\left(\frac{1}{2\widetilde{\chi}_{\|}}(1-\frac{m^{2}}{m_e^2})m\right)
\label{koopmanstoLLB}
\end{equation}

Comparing this to the LLB equation with longitudinal relaxation only
and without anisotropy and external fields, we can write Eq. (\ref{koopmanstoLLB}) in terms of $\mathbf{n}$:
\begin{equation}
\frac{dn}{dt}=\gamma\frac{\lambda}{m_e}\frac{2T_e}{3T_{C}}\frac{2q}{\sinh\left(2q\right)}H_{\mathrm{eff}}=\gamma\alpha_{\parallel}H_{\mathrm{eff}}
\label{eq:LLBappendix}
\end{equation}
where $H_{\mathrm{eff}}=\frac{m_e}{2\widetilde{\chi}_{\|}}(1-n^{2})n$, and
\begin{equation}
 \alpha_{\parallel}=\left[\frac{3R}{2\gamma}\frac{\mu_{\text{at}}}{k_{B}T_{C}}\frac{T_{p}}{T_{e}}\right]\frac{2T_e}{3T_{C}}\frac{2q}{\sinh\left(2q\right)}
\end{equation}
Thus the Koopmans' M3TM equation is equivalent to the LLB equation
with $S=1/2$ and where the precessional aspects
are not considered. The link between both of them is the identification
\begin{equation}
\lambda=\frac{3R}{2\gamma}\frac{\mu_{\text{at}}}{k_{B}T_{C}}\frac{T_{p}}{T_{e}}
\end{equation}
 As an example we  compare the result of the longitudinal relaxation in a numerical
experiment for both M3TM
and  LLB ($S=1/2$) equations. The system is put in a saturated
state with $S_{z}/S=1$ and we let it relax towards the equilibrium state.
The comparison of the results  for the temperature $T/T_{C}=0.8$ are presented in Fig.\ref{compare}.


\end{document}